\documentclass[10pt,a4paper]{article}
\usepackage{amsmath}
\usepackage{amssymb}
\usepackage{amsfonts}
\usepackage{epsfig}
\newcommand{\la}{\langle}
\newcommand{\ra}{\rangle}
\newcommand{\ovl}[1]{\overline{#1}}
\newcommand{\avg}[1]{\langle{#1}\rangle}

\newcommand{\req}[1]{(\ref{#1})}

\def\sign{\hbox{sign}\,}
\def\Tr{\hbox{Tr}}

\title{Replica symmetry breaking in the minority game}
\author{Andrea De Martino\footnote{E-mail address:
andemar@sissa.it}~
and Matteo Marsili\footnote{E-mail address: marsili@sissa.it}\\
{\normalsize \emph{International School for Advanced Studies 
(\textsc{Sissa/Isas})}}\\
{\normalsize \emph{and Istituto Nazionale per la Fisica della Materia 
(\textsc{Infm})}}\\
{\normalsize \emph{Via Beirut 2-4, Trieste, Italy}}}
\date{}
\begin{document}
\maketitle
\noindent
{\small \textbf{Abstract -}}
{\small 
We extend and complete recent work concerning the analytic solution of the
minority game. Nash equilibria (NE) of the game have been found to
be related to the ground states of a disordered hamiltonian
with replica symmetry breaking (RSB), signalling the presence of a 
large number of them. Here we study the number of NE both
analytically and numerically. We then analyze the stability
of the recently-obtained replica-symmetric solution and, in
the region where it becomes unstable, derive the solution within
one-step RSB approximation. We are finally able to draw a detailed 
phase diagram of the model. 
}\\
\mbox{}
\newline
\\
\section{Introduction}

The minority game has drawn much attention recently as a toy model of
a market \cite{hp,cz1,zenews}. In the simplest possible case, when no
public information \cite{trading} is present, its definition
is fairly simple. At 
each time step, $N$ players have to choose between two actions, such
as buying a certain stock or selling it. Those who end up in the
minority side win. This mechanism can be obtained by abstracting the
well known law of supply-and-demand. When the majority of traders is
buying a certain asset it is convenient to be a seller, for prices are
likely to be high, and viceversa. The minority side has an advantage.

The full complexity of the model arises in the presence of public
information, which is modeled by the occurrence of one of $P$ events
representing, e.g., some political news or
a price change. In the minority game agents resort to
choice rules or information-processing devices -- called strategies
henceforth -- which suggest them whether to ``buy'' or ``sell''
given the information they have received. Then each player acts
according to the suggestion of his best performing strategy. This
mechanism allows to tackle the central problem one faces when trying
to understand the collective behaviour of systems of heterogeneous
agents interacting under strategic interdependence (as in markets),
that is, how agents react to public information (e.g., political news
or price changes) and the feedback effects that these reactions have
on public information.

Early numerical simulations \cite{cz1,savit,cz2,johnson} have shown a
remarkably rich behaviour where both cooperativity and crowd
effects \cite{johnson,johnson2} arise. Much emphasis was initially
put on the emergence of a cooperative phase in the stationary state
as compared to the reference situation of random agents -- i.e.,
agents who toss a coin to decide which action to take. 
Agents in the minority game are able to coordinate
their actions and reduce the global waste of resources below the level
corresponding to random agents.

Later work \cite{cm,cmz,mcz,cmzh} has revealed that the agents'
adaptive dynamics minimizes a global function, related to market
predictability, and that the system undergoes a phase transition
between an asymmetric and a symmetric, unpredictable market, as the
ratio $\alpha=P/N$ decreases.  A full characterization of the model's
behaviour for $N\to\infty$ was derived studying the minima of the
global function by the replica method \cite{cmz,mcz,cmzh}\footnote{ It
should be mentioned at this point that the replica approach fails to
describe the system's behaviour in a certain range of the parameters,
as discussed in Refs \cite{trading,comment}. This point will be made
more precise later in the text.}.  It was also realized that agents
can greatly improve their performance and global efficiency if they
account for their own impact on the market \cite{trading,cmz,mcz}. In
this case the steady state is a \emph{Nash equilibrium}, that is a
configuration where no agent can improve his performance by changing
his behavior if others stick to theirs. Also in this case the dynamics
is related asymptotically to the minima of a global function, and
hence statistical mechanics again allows one to describe in detail the
stationary state. However, for Nash equilibria replica symmetry
breaking (RSB) occurs. This makes the replica symmetric calculation of
Refs \cite{cmz,mcz} only an approximation.

In this paper we move the first steps towards a complete
characterisation of the set of NE of the minority game. Our analysis
will be slightly more general, for we shall be able to embody the
cooperative state as well. First, we shall briefly outline the
replica approach, showing that the replica-symmetric solution has a
limited validity due to
entropy arguments. Then we compute the number of Nash equilibria,
following Ref. \cite{bm}, and show that there are exponentially
many (in $N$).  By analogy with the de Almeida-Thouless
stability analysis of
the SK spin glass model \cite{mpv,at}, we find the phase transition
line (AT line) separating the replica symmetric from the RSB phase. Finally we
shall break the replica symmetry and study the solution in the
one-step RSB approximation (1RSB). Probably the exact solution
requires infinitely many steps of RSB, but 1RSB provides already
an extremely close agreement with numerical results. 
Finally, we will be able to draw 
complete phase diagram of the model within 1RSB. 

Our discussion will focus on the statistical mechanical properties of
the model. The economic and game-theoretic aspects of the model, which
are discussed in detail elsewhere \cite{zenews,trading,mcz,cmzh}, will
only be described briefly.

\section{The model}

\subsection{Basic definitions}

The essential ingredients of the minority game are: 
\begin{itemize}
\item $N$ players, labeled by the index $i$; 
\item for each player $i$ a strategy variable $s_i\in\{\pm 1\}$, saying
which strategy ($+1$ or $-1$) player $i$ is adopting (we restrict
ourselves to the case where players have two strategies each);
\item $P$ different information patterns, labeled by the index $\mu$. 
\end{itemize}

At each time step $t$ all players receive the same information $\mu$,
drawn at random with equal probability in $\{1,\ldots,P\}$
\cite{cavagna}. Strategies $s$ are the label of information
processing devices, that suggest an action as a value of a binary
variable (like ``buy'' or ``sell'') upon receiving information
$\mu$:
\begin{equation}
s:\{1,\ldots,P\}\ni\mu\mapsto a_{i,s}^\mu\in\{\pm 1\}.
\end{equation}
Two such strategies are assigned to each agent and they are
drawn at random and
independently for each agent, from the set of
all $2^P$ such functions. In practice, the $a_{i,s}^\mu$ play the
role of quenched disorder,
analogous to the random couplings $\{J_{ij}\}$ in spin glass models.

It is convenient to make the dependence of $a_{i,s}^\mu$ on $s$
explicit by introducing auxiliary random variables $\omega_i^\mu$ and
$\xi_i^\mu$ such that
\begin{equation}
a_{i,s}^\mu=\omega_i^\mu+s\xi_i^\mu.
\end{equation}
Clearly, both $\xi_i^\mu$ and $\omega_i^\mu$ take on values in
$\{0,\pm 1\}$ but they are not independent.

The payoff to player $i$ under information $\mu$ is defined as
\begin{equation} \label{utility}
u_i^\mu(s_i,s_{-i})=-a_{i,s_i}^\mu A^\mu,\qquad
A^\mu=\sum_{j=1,N}a_{j,s_j}^\mu
\end{equation}
where $s_{-i}=\{s_j\}_{j\neq i}$. It is positive whenever $i$ is in
the minority group, whence the name of the game. Moreover, players
interact with each other only through a global quantity (namely,
$A^\mu$). This feature clarifies the mean-field character of the
model. The total loss experienced by players under information $\mu$ simply
reads
\begin{equation} \label{aquadro}
-\sum_{i=1,N}u_i^\mu(s_i,s_{-i})=(A^\mu)^2,
\end{equation}
which is always positive.

\subsection{Dynamics}

A snapshot configuration of the system corresponds to a point
$\{s_i\}_{i=1}^N$ in the (pure) \emph{strategy space} $\{\pm
1\}^N$. The game is repeated and at each time step players face the
problem of choosing the strategy to follow. By assumption, each player
keeps a ``score'' $U_{i,s}(t)$ for each strategy $s=\pm 1$ and updates it
as the game proceeds. In the beginning, players set $U_{i,\pm
1}(0)=0$. Then for $t\geq 0$ scores are updated according to the map
\begin{equation}\label{evolscore}
U_{i,s}(t+1)=U_{i,s}(t)-\frac{1}{P}a_{i,s}^{\mu(t)}\left[
A^{\mu(t)}-\eta\left(a_{i,s_i(t)}^{\mu(t)}-a_{i,s}^{\mu(t)}\right)\right]
\end{equation}
where $\eta\in\mathbb{R}$ and $s_i(t)$ denotes the strategy that
player $i$ actually uses at time $t$. The term proportional to $\eta$
is introduced to model agents who account for their market impact. We
refer the reader to Ref. \cite{mcz} for a detailed discussion of this
term\footnote{In Refs. \cite{trading,cmz,mcz} the term proportional to
$\eta$ is $\eta\delta_{s,s_i(t)}/P$. This leads, however, to the same
results of the last term in Eq. (\ref{evolscore}), in the statistical
mechanics approach. The reason is that the approach of
Refs. \cite{cmz,mcz} is based on the average of the evolution equation
in the stationary state. Observing that
$a_{i,s_i(t)}^{\mu(t)}a_{i,s}^{\mu(t)}$ is $1$ when $s_i(t)=s$ and a
random sign, with zero average, otherwise, we find that the time
average of the last term of Eq. (\ref{evolscore}) is the same as the
average of $\eta(\delta_{s,s_i(t)}-1)/P$. Hence the two equations
are equivalent (apart from an irrelevant constant $-\eta/P$).}.

Let it suffice to say that with $\eta=0$ Eq. \req{evolscore} reduces
to the standard minority game dynamics: In this case agents reward
(penalize) strategies which would have prescribed a sign opposite (equal)
to that of $A^\mu$.
In doing so, agents ignore the fact that if they actually had played
those strategies, their contribution to $A^\mu$, and hence $A^\mu$
itself, could have changed. With $\eta=1$ instead, agents correctly
accounting for their contribution to $A^\mu$. Hence the reward to
strategy $s$ is really the payoff that agent $i$ would have received
had he played that strategy. The parameter $\eta$ tunes the extent 
to which agents account for their market impact. 

Following \cite{cavagna2}, we assume that the probability with which
player $i$ chooses the strategy to adopt at time $t$ depends on the
strategy's score as follows:
\begin{equation} \label{distresse}
{\rm Prob}\{s_i(t)=\pm 1\}=C\exp[\Gamma U_{i,\pm
1}(t)],
\end{equation} 
where $C$ is a normalization constant and $\Gamma>0$ is the 
learning rate. 
With this rule, the most successful strategy is more likely to be 
chosen\footnote{For $\eta=0$, our approach gives correct results
only for sufficiently large $\Gamma$, namely for $\Gamma>\Gamma_c(\alpha)$
of Ref. \cite{trading}}. 

Note that $A^\mu$ is the contribution of $N$ terms whereas the term
proportional to $\eta$ in Eq. (\ref{evolscore}) is of order one. 
One may naively argue that
the $\eta$ term is irrelevant, as $N\to\infty$. This is not so for
exactly the same reason for which the Onsager reaction term -- or cavity
field -- is relevant in mean-field spin glass theory \cite{mpv}
\footnote{While the contribution of other spins to the effective field
acting on spin $i$ have fluctuating signs, the contribution of spin
$i$ has always the sign of spin $i$.  Mean-field theory needs to be
corrected with the subtraction of the self-interaction from the
effective-field, which becomes a cavity field. Likewise the
contribution of agent $j\neq i$ to $A^\mu$ is uncorrelated with the
action of agent $i$, whereas the contribution of agent $i$ itself is
totally correlated with his action.}. Indeed Refs. \cite{cmz,mcz}
have shown that for $\eta=1$ the dynamics converges to a NE. This
means that, in a sense, players have become fully sophisticated.
For $\eta=0$, instead, agents to converge to a sub-optimal state.

We introduce the continuous variables (``soft spins'') 
$\phi_i(t)\in[-1,1]$ as
\begin{equation} \label{mi}
\phi_i=\avg{s_i} 
\end{equation}
where $\langle\cdots\rangle$ stands for the average over
the distribution of $s_i$ in the stationary state. 
The system is then described by a point $\{\phi_i\}_{i=1}^N$ in
the hypercube $[-1,1]^N$. Hence the $\phi_i$'s 
are the relevant {\em dynamical variables} and the phase space 
is $[-1,1]^N$.

The analytic study of the dynamics Eq. (\ref{evolscore}) has been
carried out in Ref. \cite{mcz}. We shall therefore omit the details and
limit ourselves to a brief outline of the results.  One can show that
the stationary states of the dynamics correspond to the minima of the
function
\begin{equation} \label{hamne}
H_\eta=\sum_{i\neq j}^{1,N}\overline{\xi_i^\mu\xi_j^\mu}\phi_i
\phi_j+2\sum_{i=1}^N\overline{\Omega^\mu\xi_i^\mu}\phi_i+\eta\sum_{i=1}^N
\overline{(\xi_i^\mu)^2}(1-\phi_i^2)+\overline{(\Omega^\mu)^2},
\end{equation} 
where $\overline{\cdots}=(1/P)\sum_{\mu=1,P}\cdots$ and
$\Omega^\mu=\sum_{i=1,N}\omega_i^\mu$. 

Analyzing the stationary states of the dynamics is then equivalent to
minimizing $H_\eta$. The limiting cases $\eta=0$ and $\eta=1$
correspond to
\begin{equation}
H_0=\overline{\la A^\mu\ra^2}\qquad\text{and}\qquad H_1=
\overline{\la(A^{\mu})^2\ra}
\end{equation}
respectively. $H_0$ (whose minima describe the
standard minority game) is related to the market's 
predictability or available information,
as explained at length in Refs. \cite{cm,cmz,mcz}. 
In fact, if $\la A^\mu\ra\neq 0$, then one can predict that the action
$a^\mu=-\sign\la A^\mu\ra$ is more likely to be successful than the other
whenever pattern $\mu$ arises. $H_1$ is the long time
total loss of players averaged over $\mu$, as is clear from 
Eq. (\ref{aquadro}). In previous works this quantity is usually denoted as
$\sigma^2$. We stress the fact that the minima of $H_1$ are
the game's NE.

It is easy to understand that \cite{cmz,mcz}:
\begin{enumerate}
\item $H_{0}$ is a positive definite quadratic form. 
Hence for
$\eta=0$ there is a unique stationary state, corresponding to the
cooperative state observed in early numerical simulations;
\item for any $\eta>0$ both the global efficiency and the individual
payoffs are sensibly improved with respect to the $\eta=0$ case;
\item for $\eta=1$ there is a large number of stationary states, i.e.,
of NE. These states have $\phi_i^2=1$, i.e., agents play pure strategies.
\end{enumerate}

Point 1. has been treated extensively in previous works. Here we shall
focus on points 2. and 3., namely on the $\eta>0$ case. 

\section{Replica approach}

\subsection{Replica-symmetric theory}

In order to minimize the function in Eq. (\ref{hamne}) 
we can resort to statistical mechanics methods, for
\begin{equation}
\min_{\{\phi_i\}_{i=1}^N\in[-1,+1]^N}H_\eta=-\lim_{\beta\rightarrow\infty}
\frac{1}{\beta}\log
Z(\beta)\equiv\lim_{\beta\rightarrow\infty}F_\eta(\beta),
\end{equation} 
where $Z(\beta)$ is the canonical partition function associated to
$H_\eta$. Further, since $H_\eta$ contains quenched disorder we need
to apply the replica formalism \cite{mpv} to analyze its ground states. 
If we let
$J=\{a_{i,s_i}^\mu\}$ denote collectively the disorder variables and
$E_J(\cdots)$ denote statistical average over $J$, the ``typical
free energy'' $F_\eta(\beta)$ can be obtained from the identity
\begin{equation}\label{trick}
E_J[\log Z(\beta)]=\lim_{n\rightarrow 0}\frac{E_J[Z^n(\beta)]-1}{n}.
\end{equation}
A long but standard computation (see appendices in Refs \cite{mcz,cmzh}) 
leads to the following expression:
\begin{eqnarray} \label{fe}
F_\eta(\beta) & = & \frac{\alpha}{2\beta n}{\rm Tr}
\Big\{\log\Big[\Big(1+\frac{\beta}{\alpha}\Big){\mathbb I}_n
+\frac{\beta}{\alpha}\hat{q}\Big]\Big\}+\frac{\alpha\beta}{2n}
\sum_{a\neq b}^{1,n}r_{ab}q_{ab}+\nonumber\\
 & & -\frac{1}{\beta n}\log\Big\{{\rm Tr}_s
\Big[\exp\Big(\frac{\alpha\beta^2}{2}\sum_{a\neq b}^{1,n}
r_{ab}q_{ab}\Big)\Big]\Big\}+\frac{\eta}{2}(1-Q_a).
\end{eqnarray} 
where ${\mathbb I}_n$ is the $n$ dimensional unit matrix,
$\hat{q}$ is the \emph{overlap matrix} with elements
$q_{ab}=\la \phi_i^a \phi_i^b\ra$ ($a,b=1,\ldots,n$; $a\neq b$),
and $Q_a=(1/N)\sum_{i=1,N}(\phi_i^a)^2$ ($a=1,\ldots,n$).
The quantities $r_{ab}$ and $R_a$ appear as Lagrange multipliers
associated to $q_{ab}$ and $Q_a$, respectively.

Imposing the replica-symmetric (RS) Ansatz ($q_{ab}=q$ for all $a\neq b$,
$Q_a=Q$ for all $a$, and similarly for $r_{ab}$ and $R_a$) one obtains
\begin{eqnarray}
F_\eta^{{\rm (RS)}}(\beta) & = & \frac{\alpha}{2}\frac{1+q}{\alpha+
\beta(Q-q)}+\frac{\alpha}{2\beta}\log\Big[1+\frac{\beta(Q-q)}{\alpha}
\Big]+\frac{\beta}{2}(RQ-rq)+\nonumber\\
 & & -\frac{1}{\beta}\Big\la\log\int_{-1}^1\exp[-\beta V(s;z)]ds
\Big\ra_z+\frac{\eta}{2}(1-Q).
\end{eqnarray}
with
\begin{equation}
V(s;z)=-\sqrt{\frac{\alpha r}{2}}zs-\frac{\alpha\beta}{4}(R-r_1)s^2.
\end{equation}
The ground state properties of $H_\eta$ in the $N\to\infty$ limit can 
now be studied by solving the
the saddle point equations (obtained by setting equal to zero the
derivatives of $F_\eta^{{\rm (RS)}}(\beta)$ with respect to $Q$, $q$,
$R$ and $r$) in the $\beta\rightarrow\infty$ limit, since
\begin{equation}
\lim_{N\to\infty}~\min_{\{\phi_i\}_{i=1}^N\in[-1,+1]^N}\frac{H_\eta}{N}=
\lim_{\beta\to\infty}\frac{F_\eta^{{\rm (RS)}}(\beta)|_{\text{s.p.}}}{N},
\end{equation}
the subscript s.p. indicating the function $F_\eta^{{\rm (RS)}}(\beta)$ 
computed at the saddle point values of $Q$, $q$, $R$ and $r$. This procedure 
yields the so-called RS solution.

For $\eta=0$ this solution is characterised by a phase transition
at $\alpha_c\simeq 0.3374\ldots$ separating a symmetric phase 
($\alpha<\alpha_c$) with $H_0=0$ from an asymmetric one ($\alpha>\alpha_c$)
with $H_0>0$. The ``spin susceptibility'' $\chi=\beta(Q-q)/\alpha$ diverges 
as $\alpha\to\alpha_c^+$. Also, it is known that
this solution is stable against replica symmetry
breaking for all values of the control parameter $\alpha$.
A detailed account of this case can be found in 
Ref. \cite{cmz}.

We expect replica symmetry to
breakdown for all $\eta>0$ at certain critical values of $\alpha$,
denoted by $\alpha_{AT}(\eta)$. In order to test this
prediction, we study the stability of the RS solution for generic
$\eta$.

We shall see that, in the $\beta\to\infty$ limit a RSB phase
arises. It is natural then to ask whether there is a critical
temperature $\beta_c$ separating a high temperature behavior from a
low temperature one. This question, even if not directly related to the
Minority Game, may be of interest in its own and can be answered at the
RS level. Without going into details, let it suffice to mention that
setting $q=0$ in the RS saddle point equations -- which is correct for
the high temperature phase -- one finds $\beta_c=0$ for all values of
$\alpha$ and $\eta$.

\subsection{Entropy of the RS solution for $\eta=1$}

In random Ising spin systems useful indications about the stability of
the RS solution are provided by the zero temperature entropy,
namely
\begin{equation}
S_\eta^{{\rm (RS)}}(\beta\rightarrow\infty)=\lim_{\beta\rightarrow
\infty}\beta^2\frac{\partial F_\eta^{{\rm (RS)}}(\beta)}{\partial\beta^2}.
\end{equation}
This quantity is non-negative due to the discreteness of the
configuration space of the model (i.e., $\{\pm 1\}^N$). If its
zero temperature limit turns out to be negative, the corresponding
solution is unstable and further steps in the approximation are
needed.

In our case, this point is more subtle than it seems. In fact,
$H_\eta$ is defined for continuous variables $\phi_i\in[-1,+1]$, 
and not for Ising (i.e., discrete) variables. The corresponding 
configuration space is not a discrete set
of points, but rather a continuum. Therefore, in principle, the zero
temperature entropy need not be non-negative.

For $\eta=1$, however, $H_\eta$ attains its minima on the corners of
the phase space\footnote{This is a consequence of the fact that $H_1$ is an
harmonic function of $\phi_i$, i.e. $\nabla^2_{\phi} H_1=0$. This implies
that extrema occurs on the corners.} 
$[-1,1]^N$, i.e., $\phi_i=\pm 1$ for all $i$.  Hence the zero
temperature entropy calculation is revealing.  Leaving details of the
computation aside, the final result is that for $\alpha>1/\pi$
\begin{equation}
S_{1}^{{\rm (RS)}}(\beta\rightarrow\infty)=\frac{\alpha}{2}
\Big[\frac{C}{\alpha+C}-\log\Big(1+\frac{C}{\alpha}\Big)\Big],~~~~
C=\frac{\alpha}{\sqrt{\pi\alpha}-1}.
\end{equation}
This is negative for all $\alpha>1/\pi$ and it diverges at 
$\alpha=1/\pi$. For $\alpha<1/\pi$ one finds $S_1=-\infty$.
This means that RSB occurs for all values of $\alpha$ at $\eta=1$, or, in
other words, that the study of NE requires RSB.

\subsection{The number of Nash equilibria}

A crucial feature of the occurrence of replica symmetry breaking is gained
from the study of the number of minima of $H_1$. We show here that an
exponentially (in $N$) large number of such minima occurs, i.e., that the game 
possesses an exponentially large number of NE. Our analytic
results will be supported by numerical investigations.

In order to compute the number of Nash equilibria (NE) we use the fact
that NE are in pure strategies$^5$, or, that at any NE $\phi_i=\pm 1$ for 
all $i$. Keeping this in mind, we start by considering that
NE satisfy the condition 
\begin{equation}
\phi_i\left[\ovl{u_i^\mu(+1,s_{-i})}-\ovl{u_i^\mu(-1,s_{-i})}\right]=
2\left[\ovl{(\xi_i^\mu)^2}-\ovl{A^\mu\xi_i^\mu}\phi_i\right] \ge 0,\qquad
\forall i.
\end{equation}
Hence an indicator function for NE (using $2\ovl{(\xi_i^\mu)^2}\simeq 1$) is
\begin{equation}
I_{NE}(\{\phi_i\})=\prod_{i=1}^N\theta\left(
1-2\ovl{A^\mu\xi_i^\mu}\phi_i\right)
\end{equation}
($I_{NE}=1$ if $\{\phi_i\}$ is a NE and $=0$ otherwise) and the
number of NE is just obtained summing over all configuration $\{\phi_i\}
\in\{\pm 1\}^N$
(an operation which we denote by $\Tr_\phi$). 
Hence ${\cal N}_{NE}=\Tr_\phi I_{NE}\{\phi_i\}$.
Following \cite{bm} we take the average over the disorder and
introduce the integral representation of the $\theta$ function. 
We arrive at\footnote{
The reader is warned that the $\Omega$ appearing here is in no relation
with the $\Omega^\mu$ introduced in Section 2.}
\begin{equation}
E_J\left({\cal N}_{NE}\right)=
\frac{N^2\alpha^3}{(2\pi)^2}
\int_{-\infty}^\infty d\gamma 
~d\Gamma 
~d\omega
~d\Omega
\exp\left[N\Sigma(\gamma,\Gamma,\omega,\Omega)\right]
\label{nNE}
\end{equation}
with
\begin{equation}
\Sigma(\alpha,\gamma,\Gamma,\omega,\Omega)=\alpha\omega\gamma+
\alpha^2\Omega\Gamma-\frac{\alpha}{2}\log
\left[(1+\gamma)^2+2\Gamma\right]+
\log\left[1+{\rm erf}\left(\frac{1-\omega}{2\sqrt{\Omega}}\right)\right]
\end{equation}
Eq. \req{nNE} is dominated by the saddle point of $\phi$, which is
attained at $\omega=1-\gamma$, $\Omega=\frac{1-\gamma}{\alpha(1+\gamma)}$
and $\Gamma=\frac{\gamma^2(1+\gamma)}{2(1-\gamma)}$, 
where $\gamma$ is the root of the equation
\begin{equation}
\frac{\gamma^2}{4}\frac{\alpha(1+\gamma)}{1-\gamma}=
\log\left(\gamma\sqrt{\frac{\alpha(1+\gamma)}{1-\gamma}}\right)-
\log\left\{\alpha\gamma^2\left[1+{\rm erf}\left(\frac{\gamma}{2}
\sqrt{\frac{\alpha(1+\gamma)}{1-\gamma}}\right)\right]\right\}.
\end{equation}
In terms of the solution $\gamma^*$ of this equation we have
\begin{equation}
\Sigma(\alpha)=\frac{\alpha\gamma^*}{2}(2-\gamma^*)-\frac{\alpha}{2}
\log\left(\frac{1+\gamma^*}{1-\gamma^*}\right)+
\log\left[1+{\rm erf}\left(\frac{\gamma^*}{2}
\sqrt{\frac{\alpha(1+\gamma^*)}{1-\gamma^*}}\right)\right].
\end{equation}

The behavior of $\Sigma$ as a function of $\alpha$ is shown in
Fig. 1.  As expected (see \cite{trading}), as $\alpha\to 0$ the number of NE
grows as $2^N$. Numerical results from exact
enumeration for $N\le 20$ are in very good agreement, which shows that 
the so-called annealed approximation used here (i.e., taking the average of 
${\cal N}_{NE}$) is sufficient and one does not need to introduce 
replicas (to compute the average of $\log {\cal N}_{NE}$) at this level.

\subsection{de Almeida-Thouless line}

A more thorough analysis can be obtained using the
de~Almeida-Thouless
(AT) protocol \cite{at}. In order to investigate the
stability of the RS ground
states of $H_\eta$ against RSB we compute the matrix of the second
derivatives of the general expression for the free energy,
Eq. (\ref{fe}), with respect to $q_{ab}$ and $r_{ab}$. The conditions
for RSB are then obtained by studying the effect of fluctuations in
the direction of RSB. This analysis results in an instability line -- i.e. a
family of points where the RS solution becomes unstable -- in the
parameter space $(\alpha,\eta)$, called the AT line and denoted
by $\alpha_{AT}(\eta)$. 
The resulting equation has to be solved
(numerically) together with the RS saddle point equations.

An outline of the calculation is reported in the appendix. We have
studied particularly the so called replicon mode, namely those
eigenvectors of the stability matrix which are symmetric under
interchange of all but two of the indices. The replicon mode is
typically responsible for the onset of the RSB instability. Points on
the AT line are found to satisfy the following stability condition:
\begin{equation}
\alpha\left[1-\eta\left(1+\frac{\beta(Q-q)}{\alpha}\right)\right]^2=1.
\end{equation}

The ``susceptibility'' $\chi\equiv\beta(Q-q)/\alpha$ remains finite as
$\beta\rightarrow\infty$, so that the zero temperature behaviour can
be safely detected. Results are reported in Fig. 2. For
$\eta=0$ replica symmetry is preserved for all $\alpha$. The point
$\alpha_c=0.3374\ldots$ where the second order phase transition occurs
in the standard MG, separates a line of first order phase transitions,
for $\alpha<\alpha_c$, from a second order line. For $\eta=0^+$ one
finds RSB for $\alpha_{AT}=1^+$. Finally, for $\eta=1$, RS is broken for
all $\alpha$.

\subsection{Replica symmetry breaking}

The one step breaking of replica permutation symmetry is expressed by
the Parisi Ansatz for the $q_{ab}$'s and the $Q_a$'s, where an
additional parameter denoted by $m$ is introduced: $Q_{a}=Q$ (all
$a$), $q_{ab}=q_1$ (all $a\neq b$ such that $|a-b|\leq m$) and
$q_{ab}=q_0$ (otherwise). The ``free energy'' is the same as in
Eq. (\ref{fe}), but this time the overlap matrix has to be
parameterised as
\begin{equation}
\hat{q}=q_0\epsilon_n\epsilon_n^T+(q_1-q_0){\mathbb I}_{
\frac{n}{m}}\otimes\epsilon_m\epsilon_m^T+(Q-q_1){\mathbb I}_n,
\end{equation}
where $\epsilon_n$ is the $n$-dimensional column vector with all 
components equal to one (so that $\epsilon_n\epsilon_n^T$ is the 
$n$ dimensional matrix with all elements equal to one) and we 
have used the standard tensor product.

We need to consider the matrix 
\begin{equation}
\hat{T}=\Big(1+\frac{\beta}{\alpha}\Big){\mathbb I}_n+
\frac{\beta}{\alpha}\hat{q},
\end{equation}
that is
\begin{equation}
\hat{T}=\Big[1+\frac{\beta}{\alpha}(Q-q_1)\Big]{\mathbb I}_n +
\frac{\beta}{\alpha}(1+q_0)\epsilon_n\epsilon_n^T+\frac{\beta}{
\alpha}(q_1-q_0){\mathbb I}_{\frac{n}{m}}\otimes\epsilon_m\epsilon_m^T.
\end{equation}
Using the identities
\begin{eqnarray}
{\mathbb I}_n & \equiv & {\mathbb I}_{\frac{n}{m}}\otimes 
{\mathbb I}_m\nonumber\\
\epsilon_n\epsilon_n^T & \equiv & \epsilon_{\frac{n}{m}}
\epsilon_{\frac{n}{m}}^T\otimes\epsilon_m\epsilon_m^T
\end{eqnarray}
we can decompose $\hat{T}$ into tensor products and write 
its determinant $|\hat{T}|$ in a straightforward way (we need 
$|\hat{T}|$ because ${\rm Tr}(\log\hat{T})=\log|\hat{T}|$). We get
\begin{eqnarray}
|\hat{T}| & = & \Big[1+\frac{\beta}{\alpha}(Q-q_1)+n
\frac{\beta}{\alpha}(1+q_0)+m\frac{\beta}{\alpha}(q_1-q_0)\Big]\times\nonumber\\
 & & \times\Big[1+\frac{\beta}{\alpha}(Q-q_1)+m
\frac{\beta}{\alpha}(q_1-q_0)\Big]^{\frac{n}{m}-1}
\Big[1+\frac{\beta}{\alpha}(Q-q_1)\Big]^{n-\frac{n}{m}}\nonumber.
\end{eqnarray}
This means that the first factor on the r.h.s. is an eigenvalue 
of $\hat{T}$ with multiplicity one, the second one is an 
eigenvalue with multiplicity $\frac{n}{m}-1$, and the third 
one is an eigenvalue with multiplicity $n-\frac{n}{m}$. 

Putting the latter formula into Eq. (\ref{fe}) and taking the
$n\rightarrow 0$ limit as requested by the replica trick, we obtain
the one step replica symmetry broken free energy $F_\eta^{{\rm
(1RSB)}}(\beta)$, whose final expression is
\begin{eqnarray}
F_\eta^{{\rm (1RSB)}}(\beta) & = & \frac{\alpha}{2}
\frac{1+q_0}{\alpha+\beta(Q-q_1)+m\beta(q_1-q_0)}+
\frac{\eta}{2}(1-Q)+\nonumber\\
 & & +\frac{\alpha}{2\beta}\log\Big[1+
\frac{m\beta(q_1-q_0)}{\alpha+\beta(Q-q_1)}\Big]+\nonumber\\
 & & +\frac{\alpha}{2\beta m}\log\Big[1+
\frac{m\beta(q_1-q_0)}{\alpha+\beta(Q-q_1)}\Big]+\\
 & & +\frac{\alpha\beta}{4}[RQ+(m-1)r_1 q_1-mr_0 q_0]+\nonumber\\
 & & -\frac{1}{m\beta}\Big\langle\log
\Big\langle\Big(\int_{-1}^{1}ds\exp[-\beta V(s;y,z)]\Big)^m
\Big\rangle_y\Big\rangle_z\nonumber,
\end{eqnarray}
where $\langle\cdots\rangle_{x}$ denotes again the average over the unit
gaussian variables $x$ and 
\begin{equation}
V(s;y,z)=-\sqrt{\frac{\alpha r_0}{2}}zs-\sqrt{\frac{\alpha(r_1-r_0)}{2}}
ys-\frac{\alpha\beta}{4}(R-r_1)s^2.
\end{equation}

$F_\eta^{{\rm (1RSB)}}(\beta)$ depends on seven parameters: the three
overlap matrix elements $Q$, 
$q_0$, $q_1$,
their related Lagrange multipliers $R$, $r_0$ and $r_1$, and 
$m$. Their values have to be determined self-consistently from the seven
saddle point equations obtained by setting to
zero the derivatives of the free energy with respect to the above
parameters. These equations can be solved numerically.

One finds three different regimes in the $(\alpha,\eta)$ plane
when $\beta\to\infty$ (Figure 2):
\begin{enumerate}
\item For $\alpha<\alpha_0\simeq 0.09012\ldots$ (all $\eta>0$)
one has $H_\eta=0$. 
The solution does
not depend on $\eta$ as long as $\eta>0$. The self-overlap is $Q=1$
signalling that agents play pure strategies ($\phi_i=\pm 1$) but off
diagonal overlaps $q_1>q_0$ are both less than $1$. This suggests that 
NE are organized in a complex geometric structure. The parameter $m$ 
attains a finite value.
\item For $\alpha_0<\alpha<\alpha_1(\eta)$ 
(all $\eta>0$) the solution has
$H_\eta>0$, it is independent of $\eta$ (for $\eta>0$) and $1=Q=q_1>q_0$. 
The spin susceptibility $\chi=\beta(Q-q_1)/\alpha$
attains a finite value in the limit $\beta\to \infty$, which diverges as
$\alpha\to\alpha_0$. Again agents
play pure strategies and $q_0<1$ is the typical overlap between two
NE.  The parameter $m$
vanishes as $1/\beta$ (indeed the $\beta m$ is
finite as $\beta\to \infty$). The line $\alpha_1(\eta)$ is determined by the
solution of
\begin{equation}
\frac{\eta}{2}=\frac{1}{\alpha+\beta(Q-q_1)}.
\end{equation}
\item In between the line $\alpha_1(\eta)$ and the stability line
$\alpha_{\text{AT}}(\eta)$ the solution has $H_\eta>0$ and $1>Q=q_1>q_0$. 
Hence agents do not play pure strategies. The solution in this region
depends on $\eta$.
\end{enumerate}

We stress again that NE are in pure strategies, since at $\eta=1$ for
all values of $\alpha$ one finds $Q=1$. 

Figure 3 shows that the one-step calculation for $H_1/N$ agrees
very well with numerical simulations and it represents a considerable
improvement over the replica symmetric result\footnote{Note that
numerical results refer to a typical NE which need not be the ground
state of $H_1$.} Further steps of RSB, most probably infinitely many, are
likely to be needed to recover exact results. However, already the one
step calculation provides a rather good approximation.

\section{Conclusion}

Summarizing, we have analyzed the solution of the minority game by
means of statistical mechanics methods. Our starting point has been
the study of Refs \cite{cmz,mcz}, where the NE of the game have been
mapped onto the ground states of a disordered hamiltonian with RSB,
suggesting the existence of a very large number of them. First, we
have computed (both analitically and numerically) the number of NE,
showing that actually they are exponentially many in $N$ (number of
players). Then, we probed the stability of the replica symmetric
theory developed in Ref. \cite{cmz}. After showing the necessity of
RSB by simple entropy considerations, we have calculated the
instability line (AT line) using the de~Almeida-Thouless method.
Finally, we have derived the broken-replica-symmetry solution, drawing
the complete phase diagram of the model. All our results are in
excellent agreement with computer experiments.

To our knowledge, the minority game is the first example of a market
game that requires the full use of spin glass theory in order to
uncover its behaviour. Remarkably, many features
actually observed in real markets can be recovered within the simple
setup of the minority game \cite{zenews,trading,savit}.  Understanding
real markets is among the most challenging theoretical problems ahead
of us. This work suggests that statistical mechanics of disordered 
systems may be a valuable tool in this endeavour.

\textbf{Acknowledgments} We acknowledge frequent discussions with 
D. Challet, S. Franz, L. Giada,
F. Ricci-Tersenghi, R. Zecchina and Y.C. Zhang which helped us
considerably in the developement of this work.

\appendix

\section{Calculation of the AT line}

The stability matrix has dimension $n(n-1)\times n(n-1)$ and is given by
\begin{equation}
C=
\begin{pmatrix}
A^{(ab,cd)} & D^{(ab,cd)} \\ D^{(ab,cd)} & B^{(ab,cd)}
\end{pmatrix}
\end{equation}
where 
\begin{equation}
A^{(ab,cd)}=\frac{\partial^2(nF)}{\partial q_{ab}\partial q_{cd}},
\qquad B^{(ab,cd)}=\frac{\partial^2(nF)}{\partial r_{ab}\partial 
r_{cd}}\qquad{\rm and}\qquad D^{(ab,cd)}=
\frac{\partial^2(nF)}{\partial q_{ab}\partial r_{cd}}
\end{equation}
($F$ denotes shortly the replica-symmetric free energy.)

Introducing the ``perturbation'' of the RS solution in the form 
\begin{equation}
\delta q_{ab}=\zeta_{ab}\qquad{\rm and}\qquad\delta r_{ab}=x\zeta_{ab}
\end{equation}
with the condition $\sum_{b}\zeta_{ab}=0$ for all $a$, it is 
possible to show that this condition is satisfied for all $n$ by
\begin{eqnarray}
\zeta_{ab}=\zeta &(a,b)\neq(1,2)\nonumber\\
\zeta_{1b}=\zeta_{2b}=\frac{1}{2}(3-n)\zeta &b\neq 1,2\nonumber\\
\zeta_{12}=\frac{1}{2}(2-n)(3-n)\zeta\qquad &(a,b)=(1,2)\\
\zeta_{aa}=0 &\forall a.\nonumber
\end{eqnarray}
The relevant eigenvalue equations (the so-called \emph{replicon mode}) 
are given by
\begin{eqnarray}
\sum_{cd}\big(A^{(ab,cd)}+x D^{(ab,cd)}\big)\zeta_{cd} & = & \lambda
\zeta_{ab}\nonumber\\
\sum_{cd}\big(x A^{(ab,cd)}+D^{(ab,cd)}\big)\zeta_{cd} & = & \lambda 
x \zeta_{ab}
\end{eqnarray}
One needs to find an expression for the matrix elemnts $A$, $B$ and $D$.

There are three different types of matrix elements in the replica 
symmetric state, corresponding to the cases $(a,b)=(c,d)$, $a=c$ 
and $(a,b)\neq (c,d)$ respectively. For the $A^{(ab,cd)}$ they 
are $A^{(ab,ab)}$, $A^{(ab,ad)}$ and $A^{(ab,cd)}$. It is simple 
to show that in the replica symmetric state
\begin{eqnarray}
A^{(ab,ab)} & = & -\alpha\beta(E_{ab^2}+E_{aa}^2)\nonumber\\
A^{(ab,ad)} & = & -\alpha\beta[E_{ab}(E_{ab}+E_{aa})]\nonumber\\
A^{(ab,cd)} & = & -2\alpha\beta E_{ab}^2
\end{eqnarray}
where
\begin{equation}
E_{ab}=\beta q [\alpha+\beta(Q-q)]^{-2}\qquad{\rm and}\qquad 
E_{aa}=E_{ab}+[\alpha+\beta(Q-q)]^{-1}.
\end{equation}

For the $B^{(ab,cd)}$ we find
\begin{eqnarray}
B^{(ab,ab)} & = & -\alpha^2\beta^3\big(\langle\langle s^2\rangle^2
\rangle_z-\langle\langle s\rangle^2\rangle^2_z\big)\nonumber\\
B^{(ab,ad)} & = & -\alpha^2\beta^3\big(\langle\langle s^2\rangle
\langle s\rangle^2\rangle_z-\langle\langle s\rangle^2\rangle_z^2\big)\\
B^{(ab,cd)} & = & -\alpha^2\beta^3\big(\langle\langle s\rangle^4
\rangle_z-\langle\langle s\rangle^2\rangle_z^2\big)\nonumber
\end{eqnarray} 

As for $D^{(ab,cd)}$, one finds the general result
\begin{equation}
D^{(ab,cd)}=\alpha\beta\big(\delta_{ac}\delta_{bd}+\delta_{ad}
\delta_{bc}\big).
\end{equation}

It is important to notice that the relevant combinations of matrix 
elements appearing in the eigenvalue equations are of the form 
$A^{(ab,cd)}-2 A^{(ab,ad)}+A^{(ab,cd)}$, and that the eigenvalues 
can be shown to depend only on
\begin{equation}
a=A^{(ab,cd)}-2 A^{(ab,ad)}+A^{(ab,cd)}\qquad{\rm and}\qquad 
b=B^{(ab,cd)}-2 B^{(ab,ad)}+B^{(ab,cd)}
\end{equation}
via the simple formula
\begin{equation}
\lambda_{\pm}=\frac{1}{2}\Big(a+b\pm\sqrt{(a-b)^2+4}\Big).
\end{equation}

Putting things together and solving the eigenvalue equations, 
one finds that one of the eigenvalue (namely $\lambda_-$) is 
constant in sign (at least at low temperatures). The second one, 
instead, changes sign and signals the onset of RSB instability. 
The equation corresponding to $\lambda_+=0$ in the end reads
\begin{equation}
\frac{\alpha\beta^2}{\alpha^2[1+\beta(Q-q)/\alpha]^2}
\langle(\la s^2\ra-\la s\ra^2)^2\rangle_z=1.
\end{equation}
Calculating explicitly the averages appearing in the above 
formula one arrives at the AT line reported in the text:
\begin{equation}
\alpha[1-\eta(1+\beta(Q-q)/\alpha)]^2=1.
\end{equation}

\newpage
\begin{figure}[h]
\begin{center}
\epsfig{file=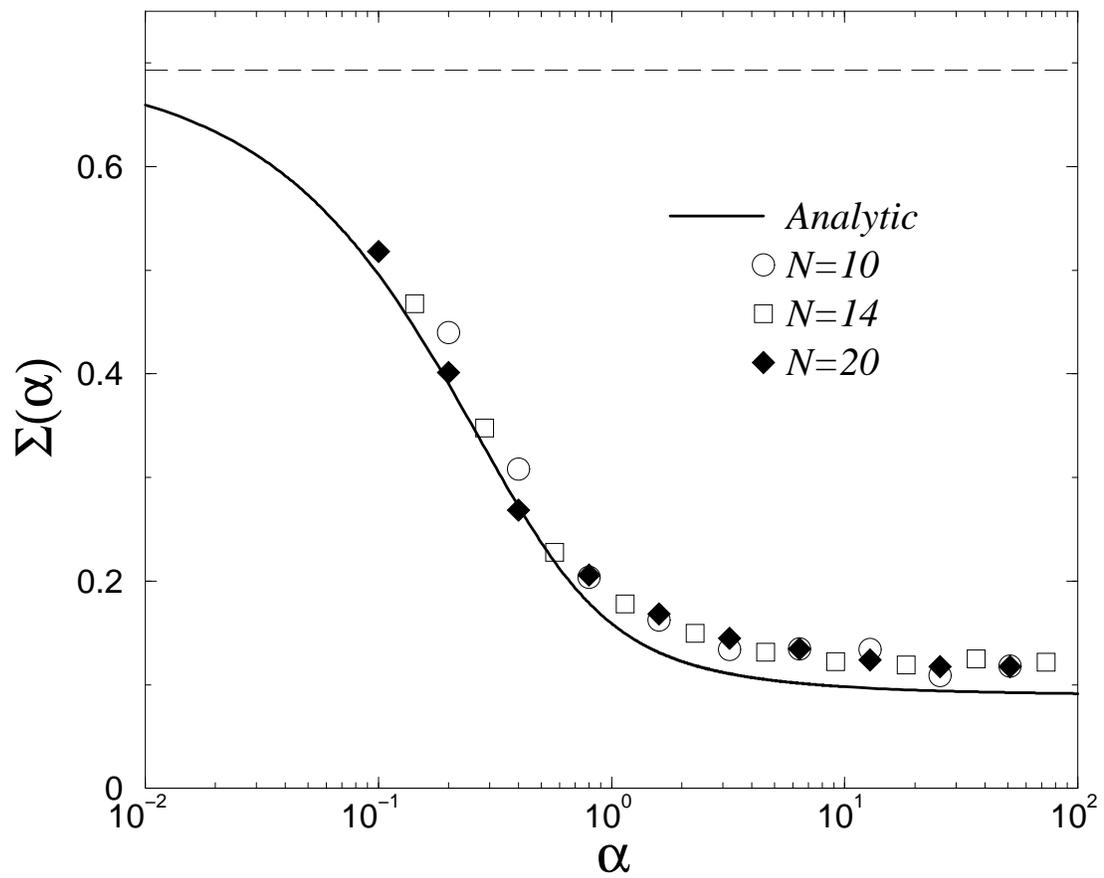,height=12cm,angle=0}
\caption{Logarithm of the average number of NE divided by $N$ ($\Sigma$)
as a function of $\alpha$.}
\end{center}
\label{nasheq}
\end{figure}

\newpage
\begin{figure}[h]
\begin{center}
\epsfig{file=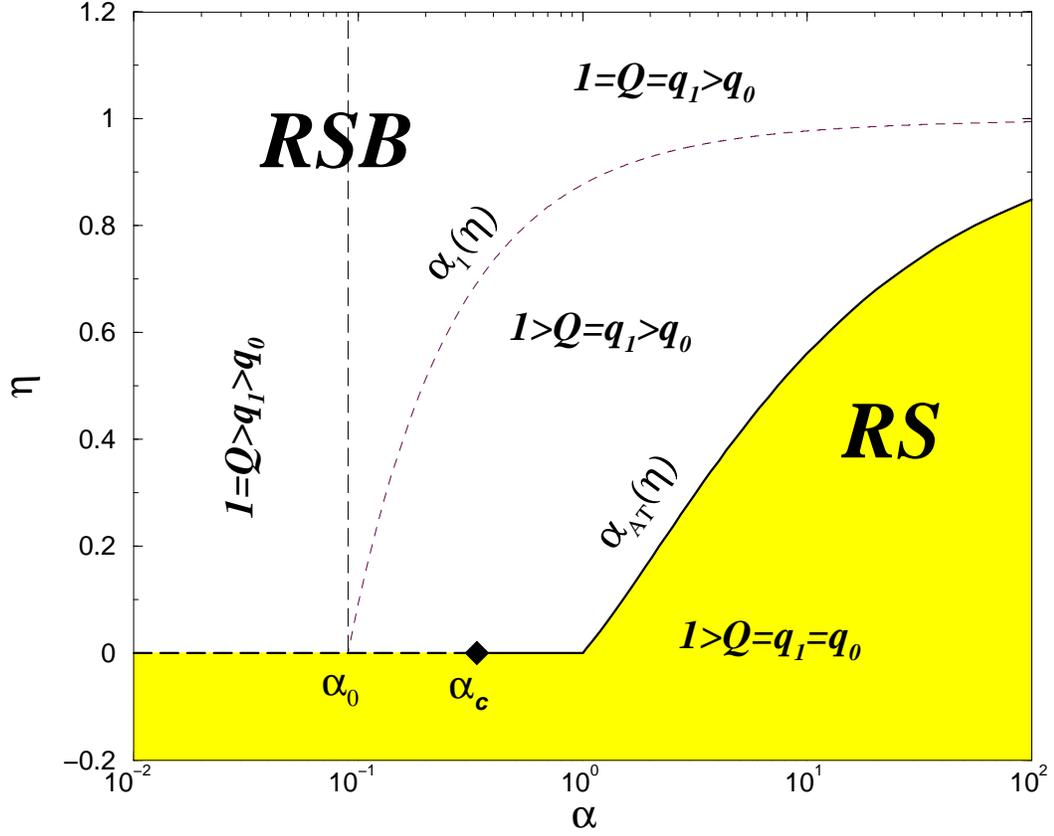,height=14cm,angle=270}
\caption{Phase diagram of the Minority Game, where the shaded region 
corrsponds to the replica-symmetric phase and the light region to the
RSB phase. 
The AT line marking the boundary between the RS and RSB
phases is denoted by $\alpha_{\text{AT}}(\eta)$.
The point $\alpha_c\simeq 0.3374\ldots$
separates a line of second order transitions (full heavy line for
$\alpha>\alpha_c$) from a first order line of critical points (heavy dashed 
line for $\alpha<\alpha_c$). Light dashed lines refer to the
different regimes found in the solution of the one-step approximation in
the RSB phase (see text for details).}
\end{center}
\label{stab}
\end{figure}

\newpage
\begin{figure}[h]
\begin{center}
\epsfig{file=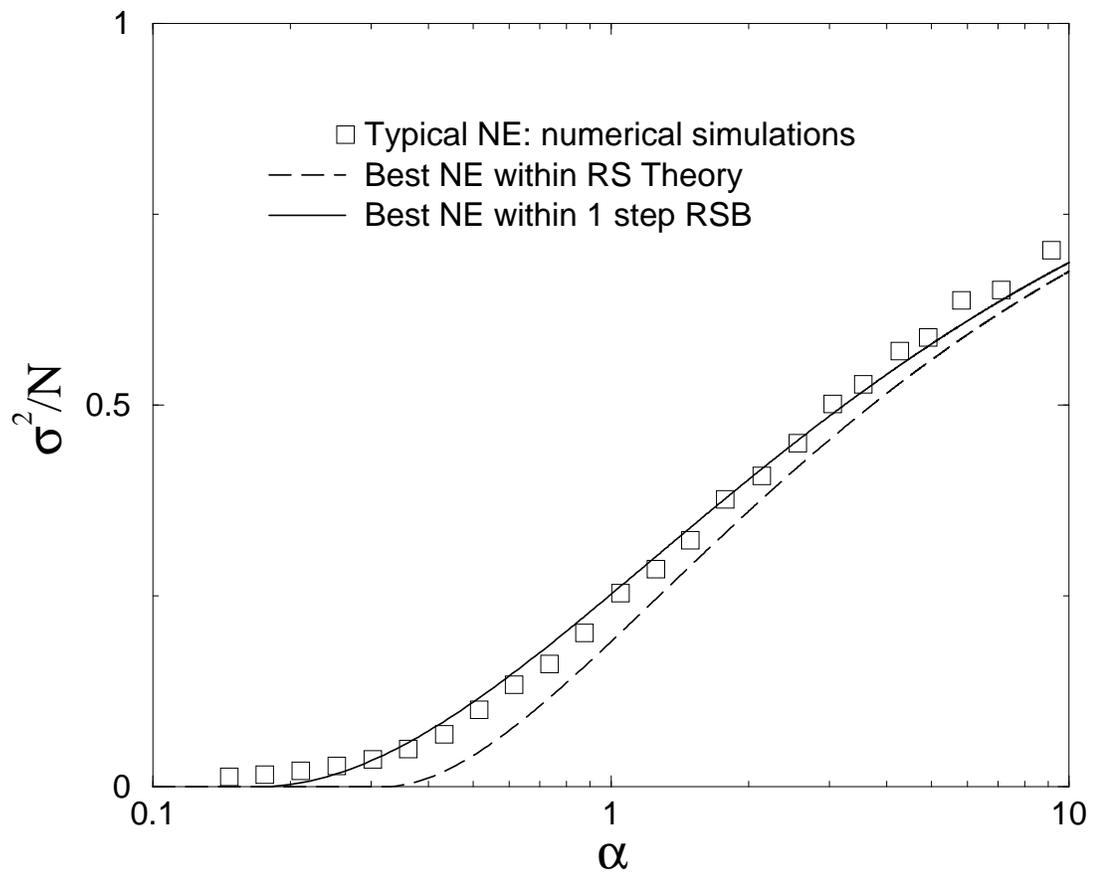,height=12cm,angle=0}
\caption{Behaviour of $\sigma^2/N\equiv H_1/N$, corresponding to NE.
The one-step solution is compared to numerical data and to the RS
solution (dashed line).}
\end{center}
\label{nash1step}
\end{figure}

\end{document}